\documentclass{aa}
\usepackage{graphicx}
\usepackage{natbib}
\usepackage{ulem}

\def\nnhp{N$_2$H$^+$}
\def\nndp{N$_2$D$^+$}
\def\hhhp{H$_3^+$}
\def\hhdp{H$_2$D$^+$}
\def\ddhp{D$_2$H$^+$}
\def\dddp{D$_3^+$}
\def\hcop{HC$^{18}$O$^+$}
\def\dcop{DCO$^+$}
\def\ammo{NH$_3$}

\def\nhh{$n$(H$_2$)}
\def\hh{H$_2$}

\def\gtsim{{_>\atop{^\sim}}}
\def\ltsim{{_<\atop{^\sim}}}

\def\tmb{$T_{\rm mb}$}
\def\tkin{$T_{\rm kin}$}
\def\txc{$T_{\rm ex}$}
\def\kms{km~s$^{-1}$}

\def\farcm{\mbox{$.\!\!^{\prime}$}}

\def\ccm{cm$^{-3}$}
\def\mic{$\mu$m}

\def\obj{LDN~1544}

\begin{document}

\title{Line profiles of molecular ions toward the pre-stellar core LDN 1544}

\author{F. F. S. van der Tak\inst{1} \and P. Caselli\inst{2} \and C. Ceccarelli\inst{3} }

\institute{
  Max-Planck-Institut f\"ur Radioastronomie, Auf dem H\"ugel 69, 53121 Bonn, Germany; \\
  \email{vdtak@mpifr-bonn.mpg.de}
  \and Osservatorio Astrofisico di Arcetri, Largo E.\ Fermi 5, 50125 Firenze, Italy
  \and Laboratoire Astrophysique de l'Observatoire de Grenoble, BP 53, 38041 Grenoble, France}

\titlerunning{Molecular line profiles toward LDN 1544}
\authorrunning{Van der Tak et al.}

\date{Received 31 January 2005 / Accepted 28 April 2005}

\abstract{Velocity profiles of ground state lines of \hhdp, \hcop\ and \nnhp, observed
  previously with the CSO and IRAM 30m telescopes, are modeled with a
  Monte Carlo radiative transfer program to study the temperature,
  density and velocity structure of the pre-stellar core LDN~1544.
  The \hhdp\ line is double-peaked like that of the other ions, but previous
  models that fit the \hcop\ and \nnhp\ profiles are found not to fit the \hhdp\ 
  data.
  Matching the \hhdp\ observations requires at least three
  modifications to the model at small radii: (1) the density profile
  must continue to rise inward and not flatten off toward the center;
  (2) the gas temperature must be nearly constant and not drop inwards
  significantly; (3) the infall velocity must increase inward, in a
  fashion intermediate between `quasi-static' (ambipolar diffusion)
  and `fully dynamic' (Larson-Penston) collapse.
  The C$^{18}$O emission indicates a chemical age of $\ltsim$0.1~Myr.
  The effects of a flattened structure and rotation on the line
  profiles are shown to be unimportant, at least on the scales probed
  by single-dish telescopes.
  Alternatively, the \hhdp\ profile is affected by absorption in
  the outer layers of the core, if gas motions in these layers are
  sufficiently small.


}
  
\maketitle

\section{Introduction}
\label{s:intro}

The formation of low-mass stars occurs in dense cores inside molecular clouds,
although the actual mechanism by which these cores accumulate material is
  still disputed (ambipolar diffusion vs.\ turbulence: \citealt{shu:ppiv};
  \citealt{hartmann01}). In any case, through the loss of turbulent and/or
magnetic support, dense cores contract towards a `critical state',
after which gravitational collapse starts and infall occurs onto a central
object, a `protostar'.
Observers distinguish such protostars in Class~0 objects, where
  the circumstellar envelope is much more massive than the star and
  accretion rates are high, and Class~I objects, where star and
  envelope have similar masses and accretion rates have dropped (see
  \citealt{andre:ppiv} for a review).
During these stages, most ($\approx$99\%) of the angular momentum of
the core is carried away by bipolar outflows, the formation of binary
stars, and magnetic fields.  The remaining 1\% leads to the formation
of the circumstellar disks commonly observed in subsequent stages of
pre-main sequence stars \citep{mundy00}.
This scenario is convincing, but many fundamental questions remain
unanswered regarding the stages before formation of a central luminous
object, the so-called pre-stellar cores. Does the collapse start from
a state of dynamical equilibrium? When does a rotating structure form,
and how important are the effects of flattening?

Detailed kinematical studies have been carried out for the Class~I
object LDN~1489 \citep{michiel01} and the Class~0 object IRAM 04191
\citep{arnaud02}.  These authors used the HCO$^+$ and CS molecules,
respectively, to trace the velocity field. 
However, at the low temperatures ($\ltsim$10~K) of pre-stellar cores,
most neutral molecules freeze out onto dust grains. In particular,
the CO and CS molecules freeze out at \nhh~$\gtsim$10$^5$~\ccm, so
that these species (as well as HCO$^+$) trace the outer parts of the
cores \citep{bacmann03,tafalla04}.
The more volatile N$_2$ molecule remains abundant at these densities,
making \nnhp\ and \nndp\ better tracers of pre-stellar core nuclei
than CO and CS (e.g., \citealt{bergin97}; \citealt{crapsi04}).
However, at the very centers of the cores, where $\gtsim$10$^6$~\ccm,
even N$_2$ freezes out, causing \nnhp\ and \nndp\ to disappear from the
gas phase \citep{bergin02,belloche04}.
One of the few molecules that do not suffer from this depletion effect
is \hhdp. This paper explores the use of \hhdp\ as kinematic tracer of
the centers of pre-stellar cores. We focus on the `prototypical'
pre-stellar core LDN 1544 in the Taurus molecular cloud ($d$=140~pc)
which has been well characterized by observations and models.
Preliminary results of this work appeared in \citet{zermatt}.

  The structure of this paper is as follows. After the data and the radiative
  program have been introduced (\S\S~\ref{sec:data}, \ref{s:mdl}), we test
  models of quasi-static collapse mediated by ambipolar diffusion, which fit
  previous observations of \obj, in spherical (\S~\ref{s:sph}) and axial
  (\S~\ref{s:disk}) symmetry. More general spherically symmetric free-fall
  models are discussed in \S~\ref{s:alto}. Sect.~\ref{s:co_dep} tests the
  adopted molecular abundance structure of the core against chemical models. The
  possibility of dynamic rather than quasi-static collapse is considered in
  \S~\ref{s:hunter}. The sensitivity of the results to the adopted collisional
  cross sections for \hhdp\ is examined in \S~\ref{s:crc}. Sect.~\ref{s:concl}
  lists the conclusions and future prospects of this work.

\section{Data}
\label{sec:data}

\begin{figure*}[tb]
\includegraphics[width=12cm,angle=-90]{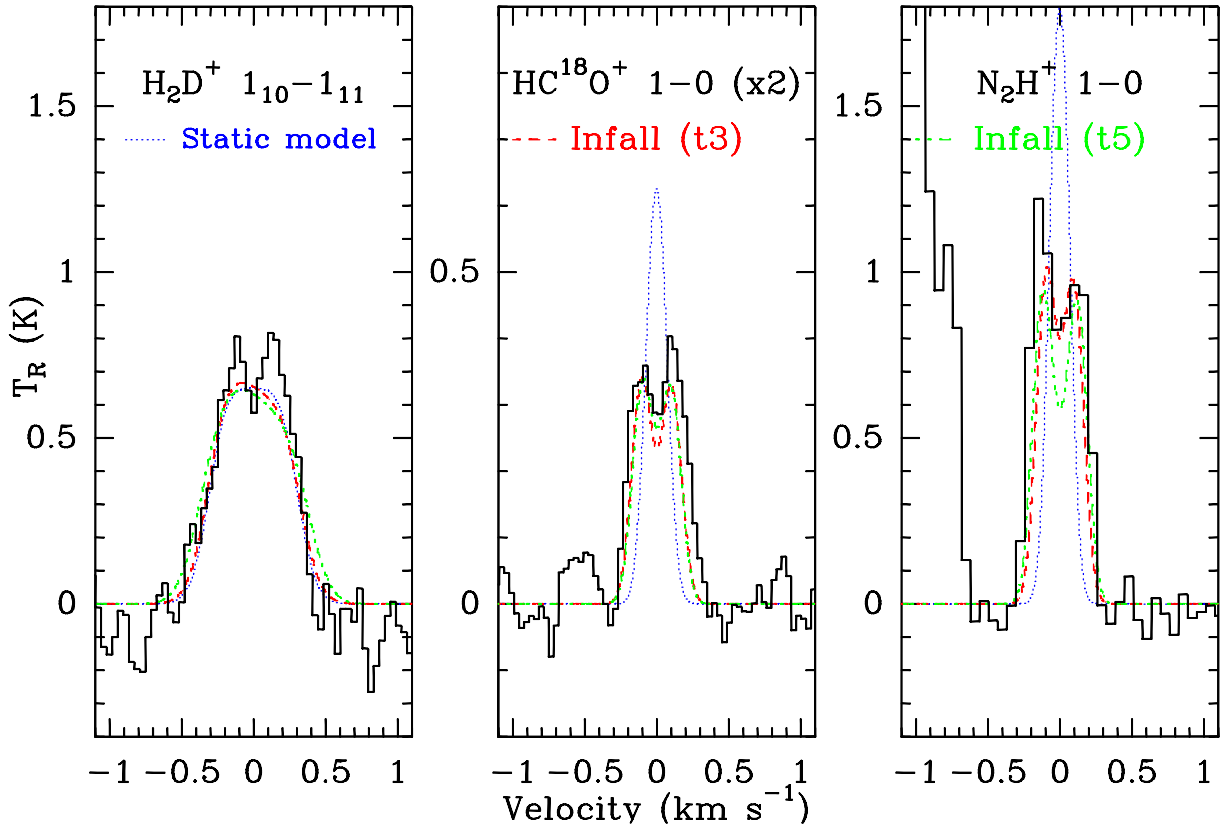}
\caption{Line profiles of \hhdp\ (left), \hcop\ (middle) and
  \nnhp\ (right; only the $F_1F$=1,0--1,1 hyperfine component)
  observed towards the `dust peak' of \obj.  Superposed are synthetic
  profiles for the best-fit static model (dotted), and infall models
  after \citet{ciolek00} for times 2.660 (dashed) and 2.684~Myr
  (dash-dotted). The line width (static: $b_D$=0.15~\kms; infall:
  $b_D$=0.05~\kms) is set to match the \hhdp\ data.}
\label{f:data}
\end{figure*}

The line profile of \hhdp\ observed toward the `dust peak' of \obj\ in
October 2002 \citep{h2d+03} suggested a double-peaked shape but
was of poor quality with low
signal to noise. Therefore the source was included in our December
2003 CSO observations, which will be presented by Caselli et al.\ (in prep.).
The left panel of Figure~\ref{f:data} shows the combined data at
the central position, which has an rms noise level of \tmb=0.11~K per
0.04~\kms\ channel.  In this spectrum, the \hhdp\ line profile is
clearly double-peaked, like the profiles of the \hcop\ and \nnhp\ 1--0
lines, observed by \citet{paola02a} and shown in the central and right
panels of the figure. To quantify the similarity between these lines,
we fitted two Gaussians to the profiles and found that each line is
well described by two thermal components at \tkin=7--10~K, separated
by $\approx$0.26~\kms\ (Table~\ref{t:twopeaks}).

\begin{table}[tb]
 \caption{Centroids and widths of the two velocity components.
Numbers in brackets denote uncertainties in units of the last decimal.}
\label{t:twopeaks}
 \centering
 \begin{tabular}{lrrr}

 \hline \noalign{\smallskip}
 \multicolumn{1}{c}{Line} & \multicolumn{1}{c}{$V_{\rm LSR}$} &
\multicolumn{1}{c}{$\Delta V_{\rm obs}$} & 
\multicolumn{1}{c}{$\Delta V_{\rm T}^a$} \cr
\multicolumn{1}{c}{} & \multicolumn{1}{c}{\kms} & 
\multicolumn{1}{c}{\kms} & \multicolumn{1}{c}{\kms} \cr
\noalign{\smallskip} \hline \noalign{\smallskip}
\hhdp\ (1$_{10}$--$1_{11}$) & 7.07(3) & 0.29(7) & 0.28--0.34 \cr
                            & 7.38(3) & 0.27(4) & \cr
\hcop\ (1--0)               & 7.04(1) & 0.18(3) & 0.10--0.12 \cr
                            & 7.28(1) & 0.23(3) & \cr 
\nnhp\ (1--0, $F_1F$=10--11)& 7.08(1) & 0.19(1) & 0.11--0.13 \cr
                            & 7.33(1) & 0.20(2) & \cr
\noalign{\smallskip} \hline \noalign{\smallskip}
\multicolumn{4}{l}{$^a$ Thermal line width at \tkin=7 and 10 K.}
\end{tabular}
\end{table}

\begin{figure*}[tb]
  \includegraphics[width=12cm,angle=-90]{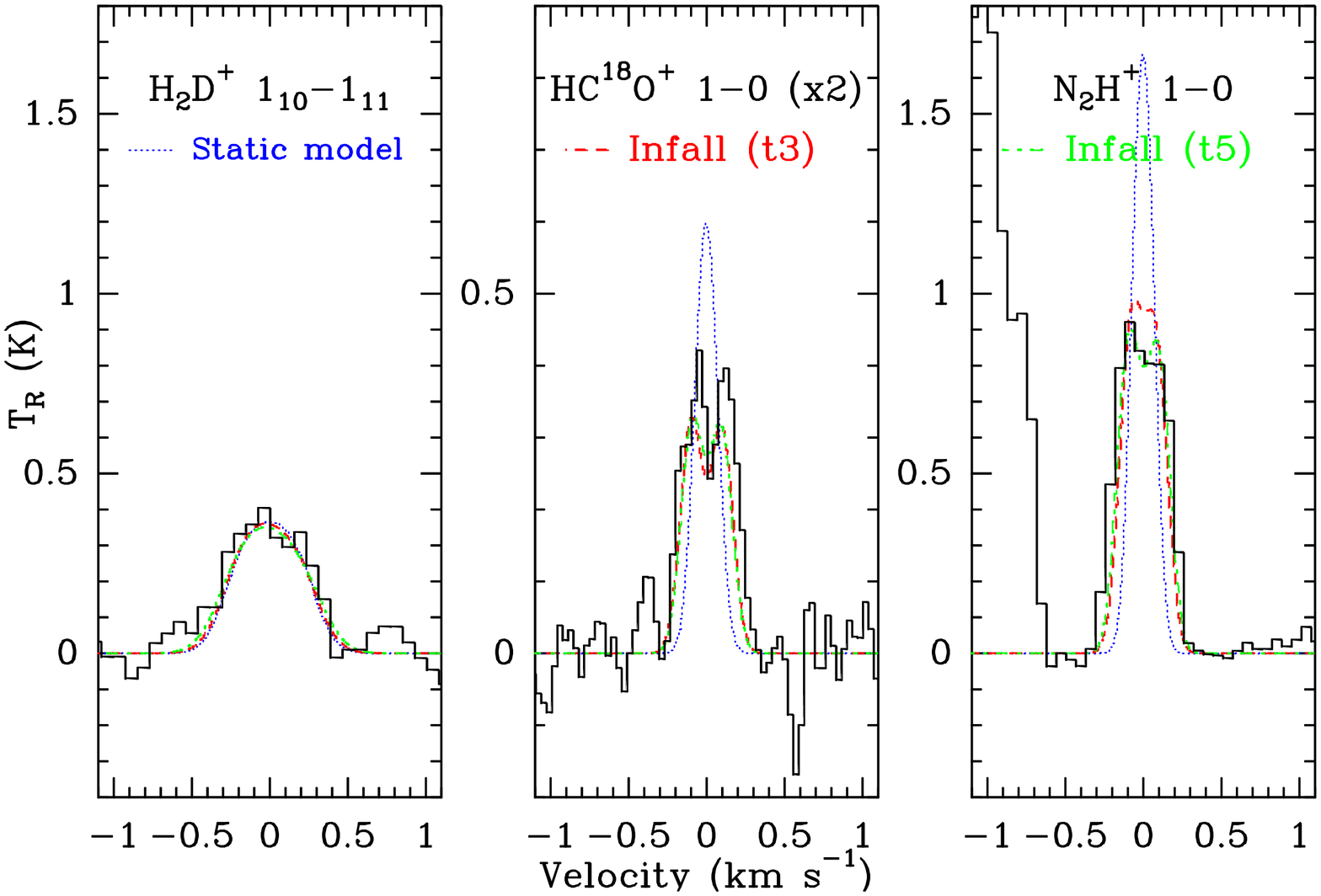}
\caption{As Figure~\ref{f:data}, at 20 arcseconds offset (average NSEW)}
\label{f:off}
\end{figure*}

The \hhdp\ line has also been detected at positions 20$''$ offset from
the dust peak, where the line is only half as strong, suggesting a
centrally peaked \hhdp\ abundance \citep{h2d+03}.
Figure~\ref{f:off} shows the line profiles of \hhdp\ and of \hcop\ and
\nnhp\ (from \citealt{paola02a}), as averages of observations taken at
20$''$ North, South, East and West offsets. For \hhdp\ and \nnhp,
neither these `average offset' spectra appear double-peaked, nor the
individual spectra.
In contrast, the average \hcop\ spectrum at the offset position
is clearly double-peaked.

The \hhdp\ line profile towards the dust peak could be affected by
absorption in the outer parts of the core, as seen in the $J$=1--0 lines of
CS and HCO$^+$ \citep{tafalla98} and \nnhp\ \citep{williams:l1544}.
Absorption is not seen for DCO$^+$ 2--1 \citep{paola02a} but this may be an
excitation effect.
Analogously, the low \hhdp\ intensity and lack of strong central absorption at
the offset position may be an excitation effect. To clarify these points,
detailed radiative transfer models have been run for \hhdp, as described in the
following sections.

\section{Model setup}
\label{s:mdl}

The line profiles of \hhdp, \hcop\ and \nnhp\ toward LDN 1544 have
been modeled with the Monte Carlo program by \citet{hst}\footnote{\tt
  http://www.mpifr-bonn.mpg.de/staff/fvandertak/ratran/},
which can treat spherical and axisymmetric geometries.
For \hcop\ and \nnhp, we use
molecular data from the database by \citet{moldata}\footnote{\tt
  http://www.strw.leidenuniv.nl/$\sim$moldata/}. For \hhdp,
collisional rate coefficients have not been calculated, so we use
scaled radiative rates \citep{black90},
which probably have an uncertainty of a factor of 3--10.
Section~\ref{s:crc} investigates the sensitivity of the
  model results to the adopted rate coefficients.

Modeling of the \nnhp\ $J$=1$\to$0 line concentrates on the lowest
frequency $F_1F$=10$\to$11 hyperfine component, which has an optical
depth of only 3.7\% of the total transition. We treat this line as if
it were an isotopic version of \nnhp: molecular excitation and line
formation are calculated with the \nnhp\ abundance reduced by 27.
Similarly, to model the \hcop\ $J$=1$\to$0 line, an oxygen isotope
ratio of $^{16}$O/$^{18}$O=500 was adopted.
We do not model H$^{13}$CO$^+$ spectra because the line profile is
complicated by hyperfine structure \citep{jsb04}.

\section{Spherically symmetric models}
\label{s:sph}

\begin{figure*}[htb]
\includegraphics[width=8cm,angle=-90]{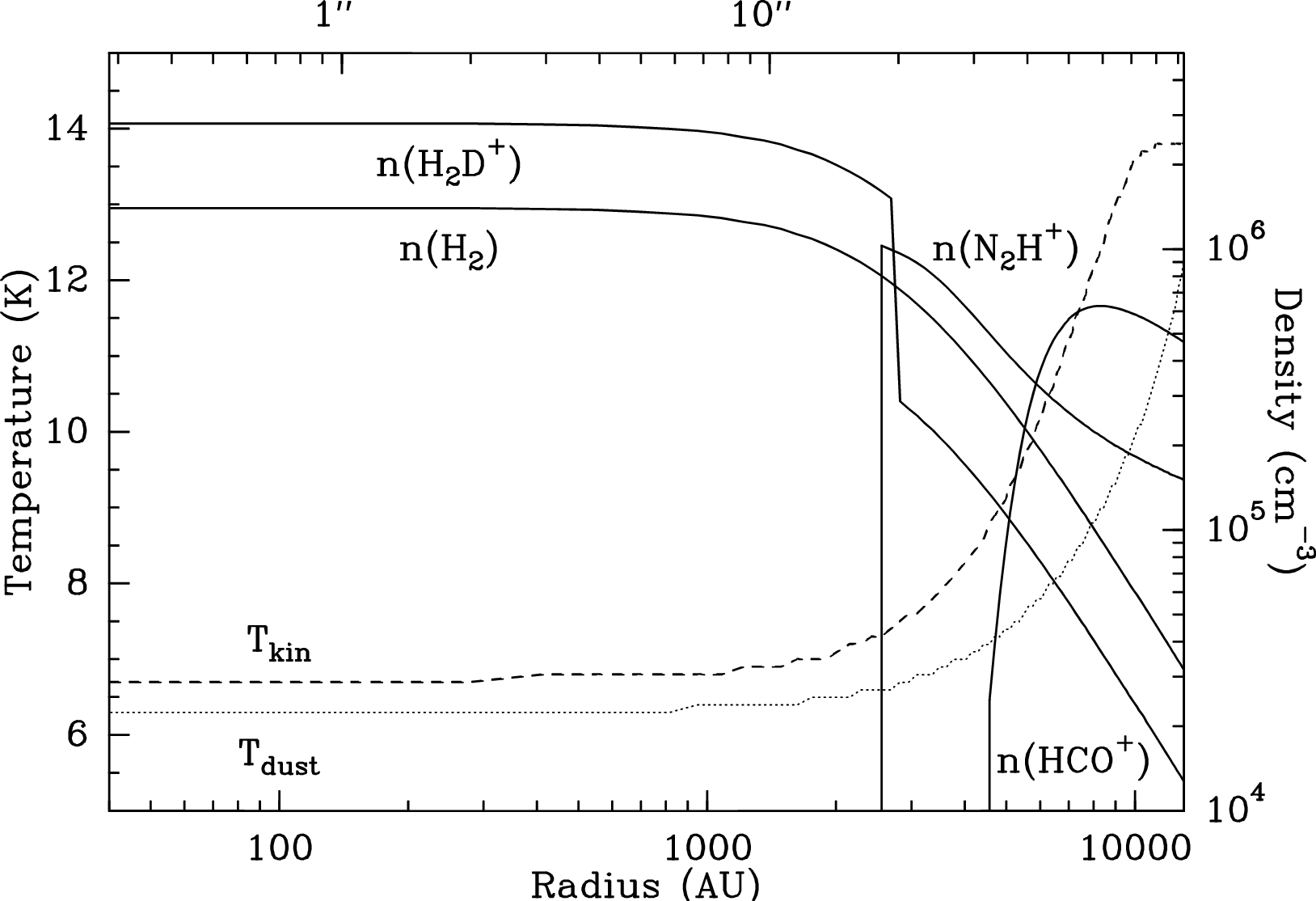}
\caption{Radial profiles of dust and gas temperature, and the
  densities of H$_2$, \hhdp\ ($\times$2$\times$10$^9$), HCO$^+$
  ($\times$10$^{12}$) and \nnhp\ ($\times$10$^{10}$) in our model of
  \obj.}
\label{f:model}
\end{figure*}

The spherical models have inner and outer radii of 80 and 13000~AU, on
a 100-point grid, logarithmically spaced.
Figure~\ref{f:model} shows the adopted temperature and density
structure, taken from \citet{galli02},
as well as the adopted molecular abundance profiles.
For HCO$^+$ and \nnhp,
we use abundance profiles from \citet{paola02b}
(Model~3 with a low ionization rate and a high sticking
  coefficient, which fits their data best), 
and assumed zero abundance inside $r$=2500~AU. 
The distribution of \hcop\ has a central hole because its chemical
precursor, CO, freezes out onto dust grains at \nhh$\gtsim$10$^5$~\ccm.
The \nnhp\ distribution continues further inward and has a hole at
smaller radii, where N$_2$ freezes out.
The assumption of an \nnhp\ `hole' of 2500~AU was already needed by
\citet{h2d+03} to reproduce the large \hhdp\ abundance at the center
of LDN~1544.
For \hhdp, we initially assumed a central abundance of
1$\times$10$^{-9}$, dropping by a factor of 5 at a radius of 20$''$
(2800~AU) \citep{zermatt} as suggested by the observations of
\citet{h2d+03}.

For the velocity field, we first adopt static models, with zero
velocity at all radii, and second models with `infall' velocity
fields. There are several competing theories for the velocity
  field of collapse onto a protostar. Here we consider models by
  \citet{ciolek00}, which assume that protostellar collapse is
  quasi-static and regulated by magnetic pressure.
Specifically, we model their simulations `t3' and `t5' which
correspond to times of 2.660 and 2.684~Myr after
the start of core collapse. At earlier times, the infall speed is
very low, while later times are implausible due to the short time
spent in these phases.
We consider these specific infall models because they are known to
  reproduce observations of HCO$^+$ and \nnhp\ toward \obj\ \citep{paola02a}.
  Sections~\ref{s:alto} and~\ref{s:hunter} treat more general collapse models.
Note that the velocity field in the simulations by Ciolek \& Basu is not
  spherically but axially symmetric with an axis ratio of $\approx$0.3
  (see their Figure 4).
Adopting these velocity fields in spherical symmetry is therefore a
simplification; however, the assumed profiles are quite similar to those of
collapsing Bonnor-Ebert spheres, as recently shown by \citet{myers05}.
Section~\ref{s:disk} will present fully consistent, two-dimensional
models with the velocity field of Ciolek \& Basu.
For the turbulent broadening, Doppler parameters $b_D$ between 0.05 and
0.25~\kms\ were tried. Smaller values of $b_D$ are overwhelmed by
thermal broadening; larger values do not fit the data.
Variations of $b_D$ with radius are considered in Section~\ref{s:crc}.

Figure~\ref{f:data} shows that the total widths of the \hhdp, \hcop\
and \nnhp\ ground-state lines can be
matched using either velocity field. The best-fit static model has
$b_D$=0.15~\kms, while $b_D$=0.05~\kms\ gives the best fits with the
infall models.  However, the observed double-peaked line shape rules
out the static model, both at the dust peak and at the 20$''$ offset
position. Using $b_D$=0.05~\kms\, the infall models can reproduce the
\hcop\ and \nnhp\ line profiles at both positions (Figure~\ref{f:off}). 
The data do not allow to decide between the `t3' and `t5' solutions.
However, none of these models reproduces the double-peaked \hhdp\ 
profile seen at the central position.

\bigskip

One reason for the lack of a central dip in the synthetic \hhdp\
spectrum may be that the model has \hhdp\ abundant at the core center,
where infall velocities are low according to \citet{ciolek00}.
Therefore we have tested the possibility that \hhdp\ has a shell-type
distribution like \hcop\ and \nnhp.
The H$_2$ molecule is too light to freeze out on dust grains and cause
depletion of \hhdp, but instead, conversion into \ddhp\ and \dddp\ in the
gas phase 
(e.g., \hhdp$+$HD$\to$\ddhp$+$H$_2$)
may cause a central \hhdp\ hole. Formation of multiply
deuterated \hhhp\ is expected at the temperature and density of the
center of \obj\ \citep{roberts03,walmsley04}.
Support for these theories comes from the recent detection of \ddhp\ 
towards the pre-stellar core LDN 1689N \citep{vastel04}.

We have run models with the \hhdp\ abundance set to zero inside inner
radii of 500--2500~AU (in steps of 500~AU), but none of these models
gives a double-peaked \hhdp\ line profile. The problem is that the
\citet{ciolek00} models predict infall velocities of $\ltsim$0.1~\kms\ 
at the radii where \hhdp\ is abundant, much less than the observed
peak separation of $\approx$0.26~\kms\ (Table~\ref{t:twopeaks}).  
Increasing the outer radius of the \hhdp\ shell from 3000~AU to 6000~AU gives
too strong line intensities at the offset position, but does produce
double-peaked \hhdp\ profiles at the center. However, the two peaks are not
$\approx$equally bright as observed, but the blueshifted peak is
$\approx$twice as strong as the redshifted one, a phenomenon known as
`infall asymmetry'.
Therefore it seems that the \hhdp\ line shape cannot be explained with
the physical and geometrical structure adopted so far.
Before changing the adopted physical structure of the core, we explore
deviations from the hitherto assumed spherical symmetry.

\section{Axisymmetric models}
\label{s:disk}

Maps of dust continuum and molecular line emission of \obj\ show a
flattened structure \citep{ward-t99,paola02a}. For example, the
1300~\mic\ dust continuum, which should be representative of the
surface density distribution, has a major axis of $\approx$2\farcm5
(21000~AU), an axis ratio of $\approx$1.7, and a position angle of
45$^\circ$ for the major axis. 
Therefore we explore the effects of flattened geometry on the line profiles.

We have constructed two-dimensional radiative transfer models with
inner and outer radii of 70 and 20000~AU, on a 30$\times$30 grid,
logarithmically spaced in both $R$ and $z$.  The density structure is
as described by \citet{ciolek00}: an `infall disk' with a flaring
angle of 15$^\circ$ and an inclination of 74$^\circ$.  
We adopt a time of 2.660~Myr, which gives the best fit to
millimeter-wave dust continuum observations \citep{ciolek00}.
The radial
infall velocity is zero at cloud center and edge, and peaks at
0.14~\kms\ at a radius of 0.03~pc.  For the gas and dust temperatures,
we tried a uniform value of 10~K (as in the model of Ciolek and Basu)
and the radial distribution by \citet{galli02} (assumed vertically
isothermal).  The abundance profiles of \nnhp, \hcop\ and \dcop\ were
again taken from \citet{paola02b}, Model~3, and for \hhdp, we used the
profile described in \S~\ref{s:sph}.

\begin{figure}[htb]
\includegraphics[width=8cm,angle=-90]{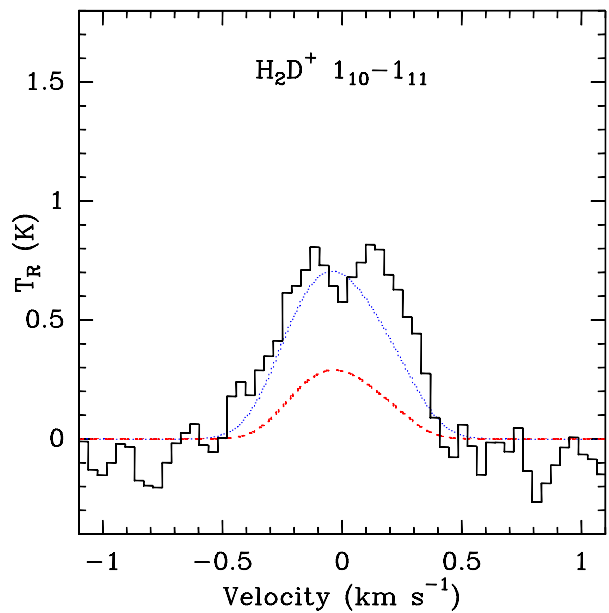}
\caption{Results of 2D models with \tkin$\equiv$10~K (dotted line) and with the
temperature distribution from Galli et al (dashed line), superposed on
the observations (histogram).}
\label{f:2d}
\end{figure}

These models reproduce the observed spatial distribution of the 
\dcop\ 2--1 and \nnhp\ 1--0 lines \citep{paola02a}, and also predict
the correct line shape for \hcop\ and \nnhp. However, they predict a
single-peaked \hhdp\ line, which is not observed (Fig.~\ref{f:2d}).

Another possible cause of the double peaked \hhdp\ line profile is
rotational motion of the central part of the core. For example,
\citet{arnaud02} found the Class~0 object IRAM 04191 to rotate
at an angular rate of $\Omega
\sim$3.9$\times$10$^{-13}$~s$^{-1}$ inside a `centrifugal radius' $r_C$ of
3500~AU. We have added rotation to the velocity field by 
\citet{ciolek00}, while keeping the density and temperature structure
from the previous models.
Rotation rates between $\Omega$=1$\times$10$^{-13}$ and
2$\times$10$^{-12}$~s$^{-1}$ were tried, inside $r_C$=3000 and
6000~AU. 
Larger values of $r_C$ are ruled out by the lack of a velocity shift
of the \hhdp\ profiles at the offset positions, and also by
the interferometric observations of \nnhp\ 1--0 by \citet{williams:l1544}.
We find that the \hhdp\ line profile is hardly changed from the
non-rotating case for $r_C$=3000~AU.
For $r_C$=6000~AU, the \hhdp\ line is considerably broadened, and the
total width of the observed \hhdp\ profile limits $\Omega$ to
$<$5$\times$10$^{-13}$~s$^{-1}$. None of these models predict
double-peaked line shapes.
In general, both flattening and rotation appear unimportant for LDN
1544 on the scales probed by single-dish observations.

\section{Alternative temperature and density structures}
\label{s:alto}

\begin{figure}[htb]
\includegraphics[width=6cm,angle=-90]{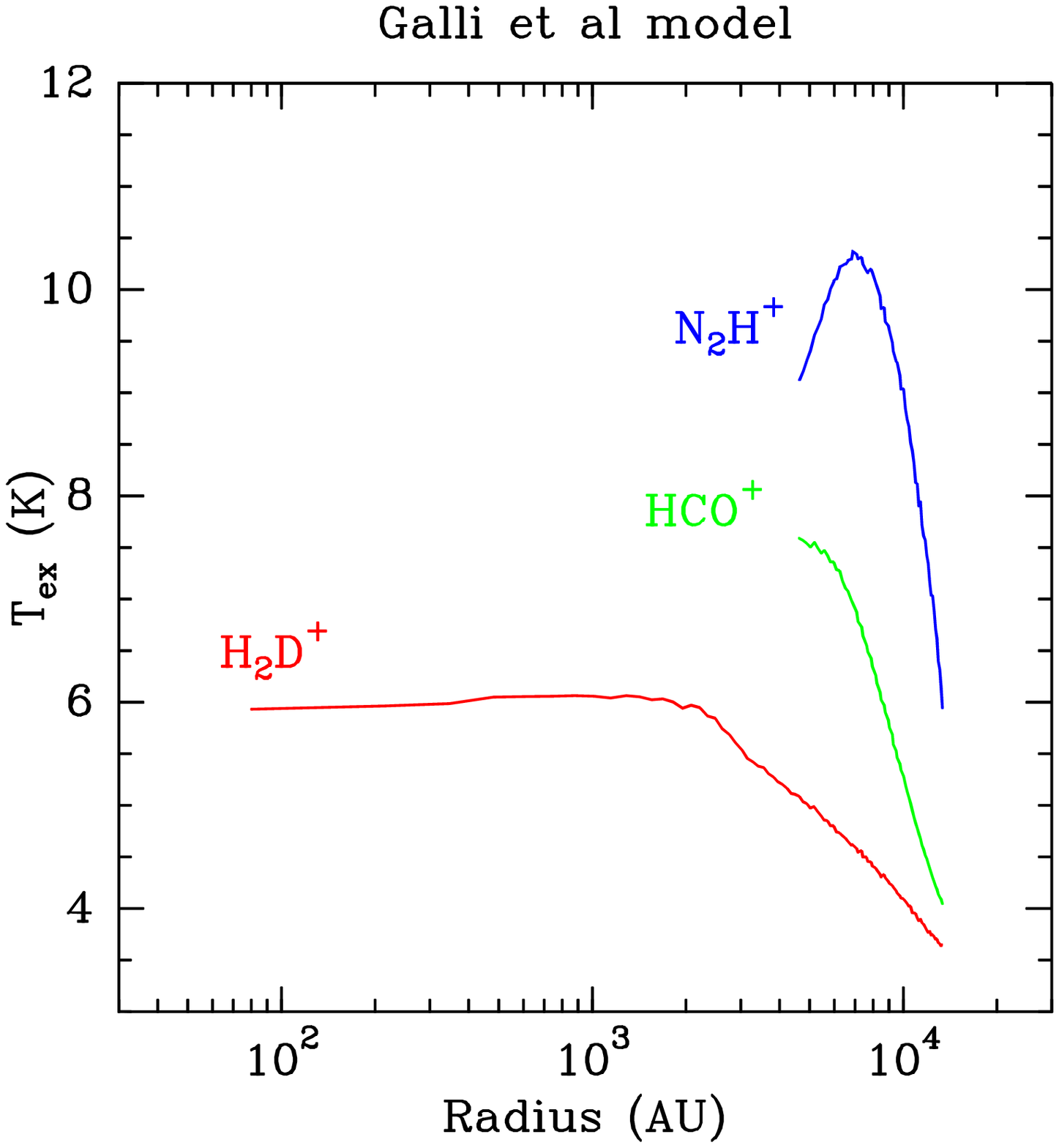}
\includegraphics[width=6cm,angle=-90]{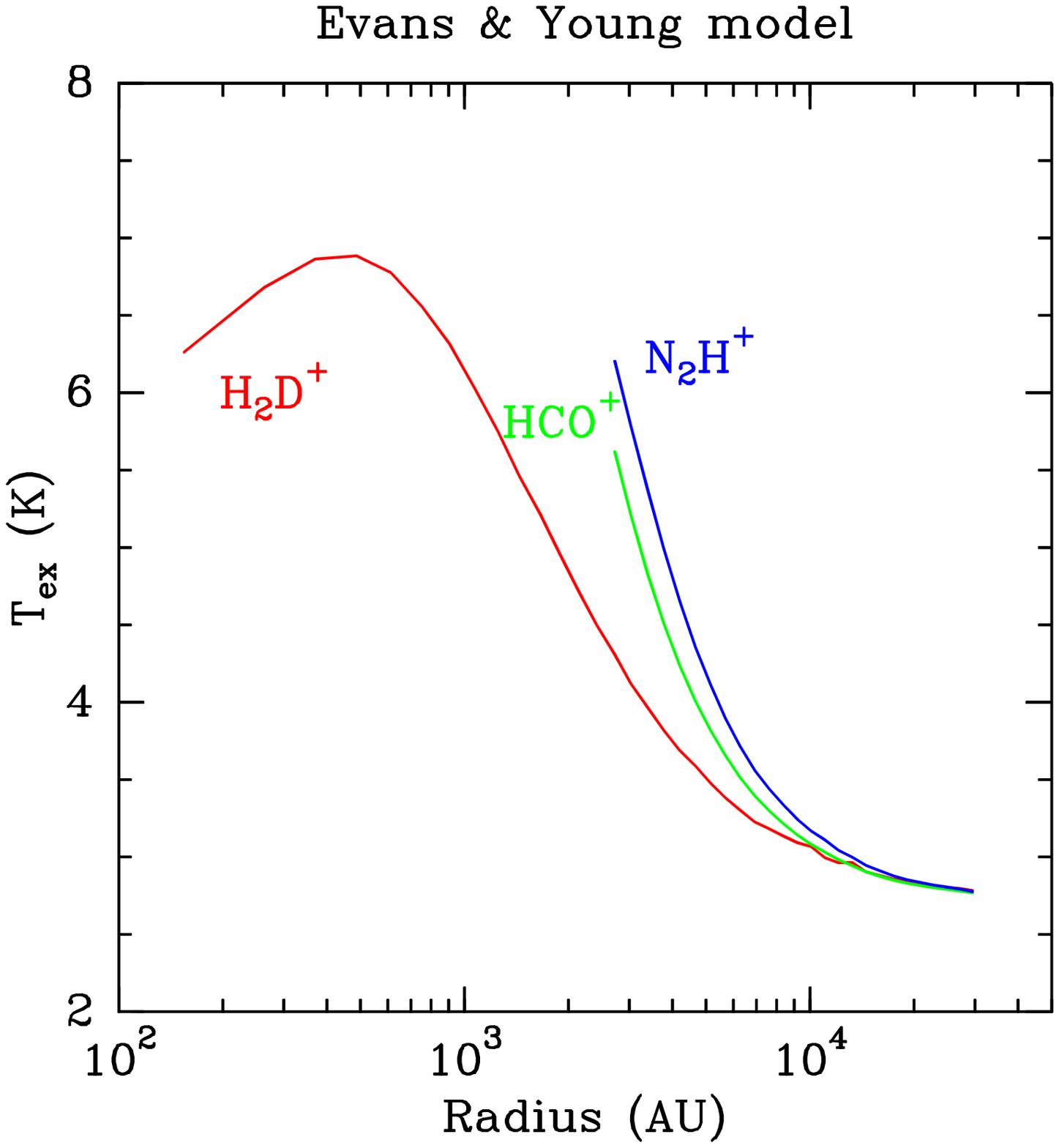}
\includegraphics[width=6cm,angle=-90]{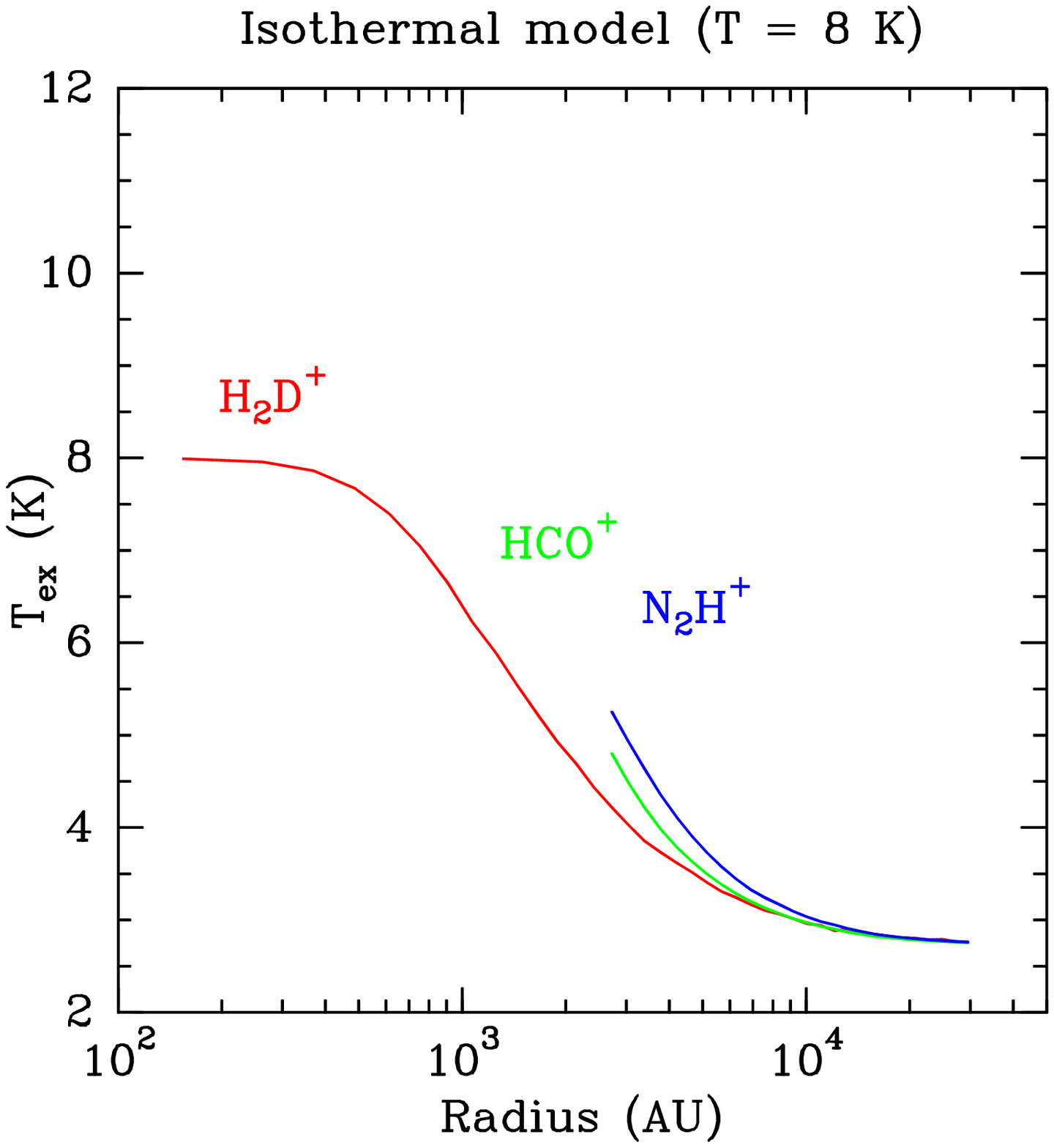}
\caption{Excitation temperature profiles of \hhdp\ $1_{10}$--$1_{11}$
  for the physical structures by Galli et al (top), Young et al
  (middle), and for the isothermal power-law model (bottom). Within the
  radius of high \hhdp\ abundance ($\sim$3000~AU), \txc\ is
  practically constant in the Galli case, but has a gradient in the
  other cases.}
\label{f:txc}
\end{figure}

The reason that the models in \S\S~\ref{s:sph} and~\ref{s:disk} do not
indicate \hhdp\ self-absorption may be that the \hhdp\ excitation
temperature is almost constant in the range of radii ($\ltsim$3000~AU)
where \hhdp\ is abundant (\hhdp/\hh~=~10$^{-9}$). 
In particular, the models from \S~\ref{s:sph} with Galli's temperature
distribution have \txc=5.8--5.6~K inside $r$=3000~AU
(Fig.~\ref{f:txc}, top), while the isothermal models of
\S~\ref{s:disk} have \txc=8.8--8.2~K at these radii.

In search of an excitation temperature gradient, we went
back to spherically symmetric models and adopted the
density and dust temperature distributions by \citet{evans01},
with gas temperatures calculated by \citet{young:h2co}. 
In these models for \obj, the density continues to rise toward the
center, rather than flattening off as in the model of \citet{ward-t99}.
As a consequence, \txc\ drops from 6.9 to 5.5~K at radii of
500-2500~AU (Fig.~\ref{f:txc}, middle).

\begin{figure}[htb]
\includegraphics[width=8cm,angle=-90]{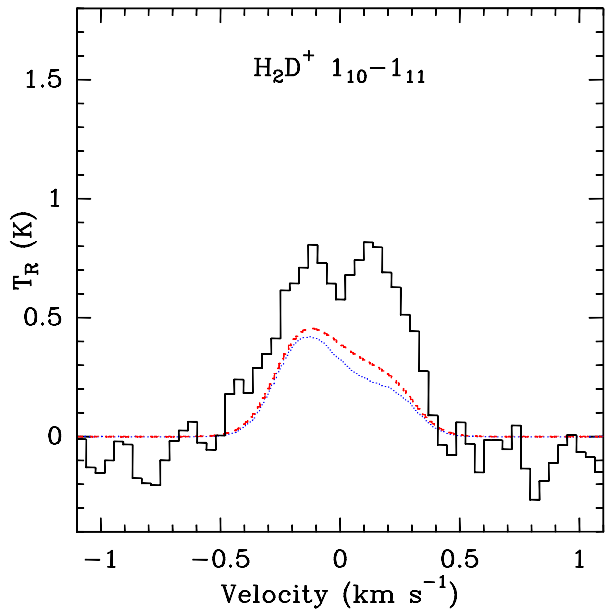}
\caption{Results of models with a power-law density distribution and
  \tkin$\equiv$8~K (dotted line) and with the gas temperature calculated by
  Young et al (dashed line), superposed on the observations (histogram).}
\label{f:alt}
\end{figure}

Combining this physical structure with velocity field `t3'
from \citet{ciolek00}, the \hhdp\ profile remains single peaked. This result
holds also when adopting constant infall speeds up to 0.2~\kms.
Larger infall speeds are inconsistent with the \hcop\ and \nnhp\ data.
Even an inward increasing velocity profile, $v=v_0(r/r_0)^{-0.5}$ with
$r_0$=10$^4$~AU and $v_0$=0.1~\kms, fails to produce a double peak
(Fig.~\ref{f:alt}).
Adopting $r_0$=10$^3$~AU and $v_0$=0.2~\kms, which mimics the `t5'
velocity field \citep{paola03} does not change this result appreciably.
Furthermore, applying central holes in the \hhdp\ abundance distribution,
with radii of 500-1500~AU, changes the total intensity of the line,
but not its shape.

To prevent \txc\ from dropping off at the very center, we took the
more drastic step of forcing a uniform \tkin, while keeping the density
power law structure by Evans et al. Values of \tkin=6~K, 8~K, 10~K and
12~K were tried. 
Fig.~\ref{f:txc} (bottom) shows the \txc\ profile for \tkin=8~K.
The \hhdp\ profiles for these models show a classic `infall asymmetry'
for abundances $\gtsim$10$^{-9}$, or a broad peak for lower abundances.
The observed profile with two peaks of equal strength is not reproduced.
Leaving the dust temperature unchanged or setting it equal to \tkin\
makes no difference to these results.

\section{Self-consistent chemical model}
\label{s:co_dep}

The rate of formation of \hhdp\ by the reaction of \hhhp\ with HD
depends mainly on the cosmic-ray ionization rate which is not expected
to change within the pre-stellar core. However, the main destruction
channel of \hhdp\ is the reaction with CO, the abundance of which is
expected to decrease significantly at low temperatures and high
densities due to freeze-out on dust grains. The models so far take
depletion into account in a simplified way by considering `jumpy'
\hhdp\ abundance profiles.
This section considers models that compute the abundances of \hhhp\ 
and its isotopomers self-consistently across the LDN~1544 core.  For
that we adapted the model of \hhhp\ chemistry in young protoplanetary
disks developed by \citet{ceccarelli05}. These models ignore the
effect of dynamics on the chemistry.
Briefly, the model computes the abundances of \hhhp, \hhdp, \ddhp\ and
\dddp\ by solving the chemical network for these species, which is
mostly governed by the CO depletion and the cosmic-ray ionization rate
(\citealt{roberts03}; \citealt{walmsley04}; see the detailed
discussion in \citealt{ceccarelli05}). 
The model uses the `fast' rate for the \hhhp\ +~HD reaction
\citep{roberts03} and 920~K for the binding energy of CO on the grain
surface \citep{bergin97}. The papers by Roberts et al.\ and
  Walmsley et al.\ discuss the uncertainties of the other relevant
  chemical reation rates such as the dissociative recombination of
  \hhdp, \ddhp\ and \dddp.
The role of cosmic rays is twofold: on the one hand, they set the
ionization degree in the gas, and therefore the total abundance of
\hhhp\ and its isotopomers in the region where CO is depleted.
The larger the ionization degree, the larger the absolute abundance
of \hhhp\ in all its isotopic forms.
Second, the cosmic rays regulate the desorption of CO molecules
from icy grain mantles back into the gas phase.
The model includes `spot heating' of grains by cosmic ray impacts, 
but no non-thermal desorption mechanisms which are much less
  effective \citep{shen04}. 
The larger the cosmic-ray flux, the higher the CO gas-phase abundance,
and the lower the abundance of the deuterated forms of \hhhp.

\begin{figure}[htb]
\includegraphics[width=8cm,angle=0]{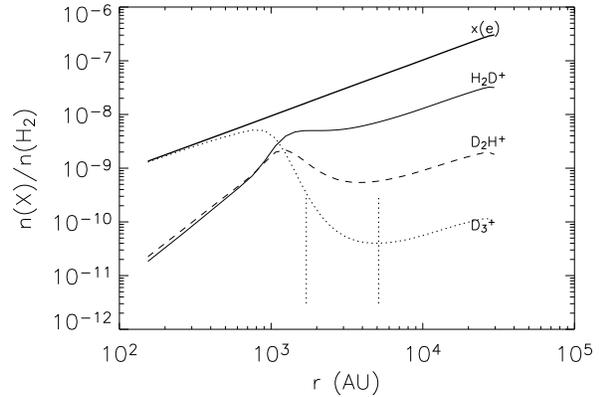}
\caption{Calculated abundances of \hhdp, \ddhp\ and \dddp\ for an age
  of 0.1~Myr, a cosmic-ray ionization rate of 3$\times$10$^{-17}$
  s$^{-1}$ and the physical structure by \citet{evans01}. The dotted
  lines indicate the radii equivalent to one and two CSO beams, for
  comparison with observations.}
\label{f:ccmod}
\end{figure}

The model assumes that at time $t$=0, all CO is in the gas phase, so
that the CO abundance is equal to the canonical value $9.5 \times
10^{-5}$ with respect to H$_2$ (e.g., \citealt{frerking82}).  As time
passes, the CO molecules condense out onto the grain mantles and
disappear from the gas phase. The time scale of the CO depletion is
set by the density and temperature, which are taken equal to those
computed in the models of Galli et al.\ and Evans et al., previously
described. 
The result is a zone where the increased CO depletion leads to an
enhanced \hhdp/\hhhp\ ratio, namely an increased \hhdp\ abundance 
(see also \citealt{h2d+03} and \citealt{ceccarelli04}).
Our computed \hhdp, \ddhp\ and \dddp\ abundances coincide with those
calculated by \citet{roberts03} and \citet{walmsley04} when the same
densities are considered.

Figure~\ref{f:ccmod} shows an example of the computed \hhdp, \ddhp\ 
and \dddp\ abundance profile across LDN~1544. Also shown is the
  electron fraction, which equals the total ionization. In these highly shielded
regions, most charge comes from molecules, not from metals, which are also
expected to be highly depleted \citep{paola02b}.
Note that in the very inner region, the \hhdp/\hhhp\ and \ddhp/\hhhp\
abundance ratios have a `plateau', whereas the \dddp/\hhhp\ abundance
ratio keeps increasing. The result is a `hole' in the \hhdp\ and
\ddhp\ abundances, because \dddp\ becomes the charge carrier when the
CO depletion is larger than $\sim$30.
The result that \dddp\ is the dominant ion at the centers of pre-stellar
  cores was found before by \citet{roberts03} and \citet{walmsley04}; adopting
  the temperature and density structure from \citet{young:h2co} for \obj, it
  happens at $R \ltsim 3000$~AU.
The \hhdp\ line profiles computed from these models depend
on two major parameters: the cosmic-ray ionization rate and the time
(for the CO depletion). We ran models varying both parameters for the
two physical structures of the core, the Galli et al.\ and Evans et
al.\ structure respectively. We considered ages of 0.05, 0.1 and
1.0~Myr, and fixed the cosmic-ray ionization rate at the canonical
value of 3$\times$10$^{-17}$ s$^{-1}$ \citep{zeta}.
The model does not consider any shielding of cosmic rays, as in
\citet{paola03}. 

Since the central CO--free zone grows in time,
the strengths of isotopic CO lines constrain the chemical age.
With the velocity field of \citet{ciolek00}, we calculate
an intensity for the C$^{18}$O 1--0 line in the 22$''$ beam of the
IRAM 30m telescope of 4.3~K for $t$=0.05~Myr, 3.5~K for 0.1~Myr and
1.5~K for 1.0~Myr. The observed strength of $\approx$5~K
\citep{paola02a} thus suggests an age of $\ltsim$0.1~Myr for LDN~1544.
It is not clear if these ages are compatible with our assumption
  of a static cloud. Models including both chemistry and dynamics
  (e.g., \citealt{rodgers:cocoon}; \citealt{aikawa05}) are needed to
  address this issue.

\begin{figure}[htb]
\includegraphics[width=8cm,angle=-90]{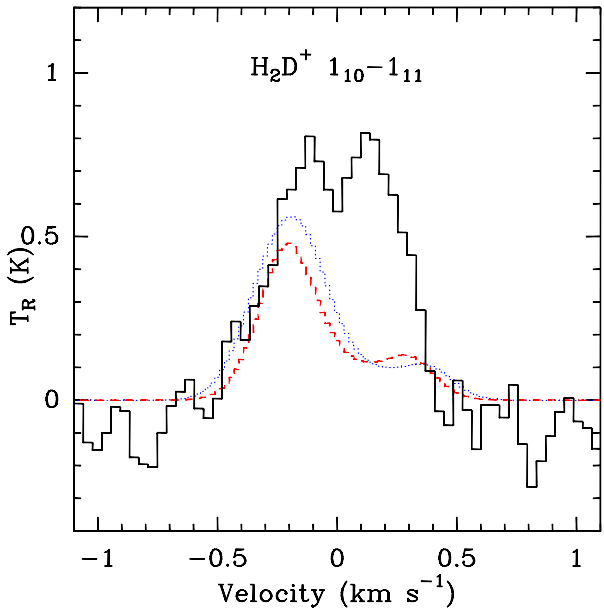}
\caption{Results of models with a power-law density distribution and the
  gas temperature calculated by Young et al, the effect of depletion
  included for a chemical age of 0.1~Myr for either a constant infall
  velocity (dotted line) or the Ciolek \& Basu velocity field (dashed
  line), superposed on the observations (histogram).}
\label{f:ccchem}
\end{figure}

To model the \hhdp\ line profile, we assume an
  ortho/para ratio of unity as appropriate for low temperatures
  and high densities, although the exact amount of molecular
  depletion may make a factor of $\sim$2 difference
  (\citealt{pagani:h2d+}; \citealt{walmsley04}).
Figure~\ref{f:ccchem} shows the \hhdp\ line profiles calculated for a
constant infall velocity of 0.2~\kms\ and for the Ciolek \& Basu
velocity field. The synthetic profiles are double peaked, but show
strong infall asymmetry while two peaks of equal strength are
observed. The strong self-absorption makes the emission
$\approx$50\% weaker than observed.  At the offset position, the
emission is also predicted to be asymmetric, and also $\approx$50\%
weaker than observed.  

These models seem to overestimate the depletion of CO and the
abundance of \hhdp. Chemical ages $<$0.05~Myr do not seem plausible,
but the age is uncertain through the dust opacity coefficient, which
makes the density uncertain by a factor of two.
Alternatively the gas temperature (which controls the emission) is
higher than the dust temperature (which controls the depletion). The
binding energy of CO to the grain surface may also have been
overestimated, so that CO does not stick to the surface as well as
assumed. 
We have not considered the reduced cosmic-ray ionization rate which
previous studies of the LDN~1544 core indicate in this source
(\citealt{paola02b}; \citealt{tafalla04}), since the CO depletion
would be even higher, implying a worse match to the data.

\section{Alternative kinematics}
\label{s:hunter}

The infall model
by \citet{ciolek00} is a magnetically mediated, `quasi-static' collapse
with relatively slow infall velocities. The collapse may instead be
`dynamic' which means that the gas never reaches equilibrium. The most
extreme case is the Larson-Penston collapse (e.g., \citealt{zhou92})
where the velocity reaches many times the sound speed. We modeled the
\hhdp\ line with such a velocity profile, but find that emission is
predicted at much higher velocities than observed.

Hydrodynamic simulations by \citet{hunter77} show the existence of a
continuum of solutions between the extreme static and dynamic cases.
The solutions for the density $\rho$ at time $t$ and radius $r$ are
$$
\frac{4\pi G\rho r^2}{a^2} = \sqrt{\frac{m_0 r}{2at}}
$$
and the velocity $u$ (directed inward) follows
$$
\frac{u}{a} = \sqrt{\frac{2m_0at}{r}}
$$
where $a$ is the sound speed, and $m_0$ is a dimensionless parameter.
The Larson-Penston collapse has $m_0 = 46.9$ while $m_0 = 0.975$ for
the quasi-static (Shu) collapse.

\begin{figure}[htb]
\includegraphics[width=8cm,angle=-90]{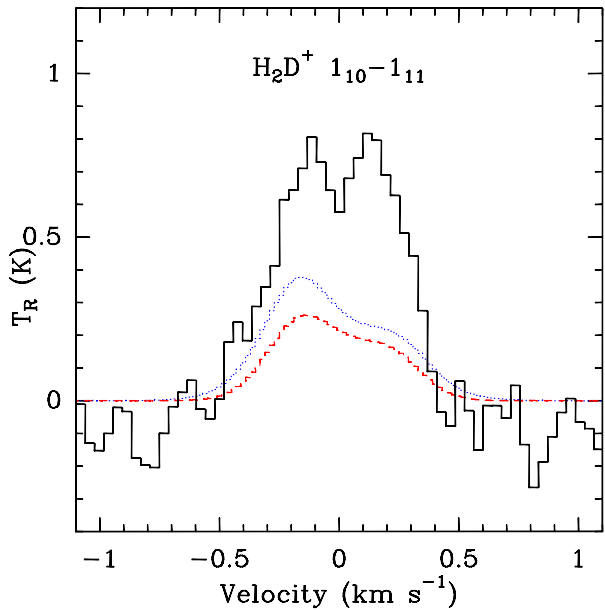}
\caption{Results of models with a power-law density distribution,
  \tkin$\equiv$10~K, Hunter's velocity profile `11d', and with the \hhdp\ 
  abundance constant (dotted line) and with an inner hole (dashed
  line), superposed on the observations (histogram).}
\label{f:hunt}
\end{figure}

As examples of infall models `intermediate' between quasi-static
and fully dynamic collapse,
we have run models for Hunter's cases 11b and 11d which have
$m_0 = 2.577$ and $m_0 = 1.138$ respectively. The gas and dust
temperatures in these models are equal at 10~K, corresponding to a
sound speed of 0.19~\kms.
Fig.~\ref{f:hunt} presents synthetic line profiles for Model 11d.
Results are shown for the case of a constant \hhdp\ abundance and of
an inner hole in the \hhdp\ distribution. These models have peaks at
the observed velocities, but with `infall asymmetry': the blueshifted
peak is stronger than the redshifted one.  The results of Model 11b
are similar, but with this asymmetry more pronounced.

\section{Effect of the collisional rate coefficient}
\label{s:crc}

Another potentially important parameter for the \hhdp\ line formation is the
collisional rate coefficient (CRC). To quantify the sensitivity of our model
results on the CRC, we have run models with the CRC increased and decreased by
factors of~3 and~10.
These models use the power-law density profile from \citet{evans01},
the temperature profile from \citet{young:h2co} and the \hhdp\
abundance profile from \citet{ceccarelli05}. 
Since the previous sections indicate that the details of the infall
velocity field play a minor role for our models, we simplify the
kinematics to a `step function' in $b_D$, the turbulent line width.
The observations suggest that $b_D$ is larger in the (emitting)
central part of the core than in the (absorbing) outer layers, so we
adopt $b_D$=0.1~\kms\ at radii $<$10$^3$~AU and $b_D$=0 outside this
radius. The model does not include any systematic radial motions, but
the increased $b_D$ at the center is presumed to arise in infall.
This setup implies the existence of infall motions, while the results
do not depend on the specific infall model. Note that thermal
broadening is added to the turbulence at all radii.

\begin{figure}[htb]
\includegraphics[width=8cm,angle=-90]{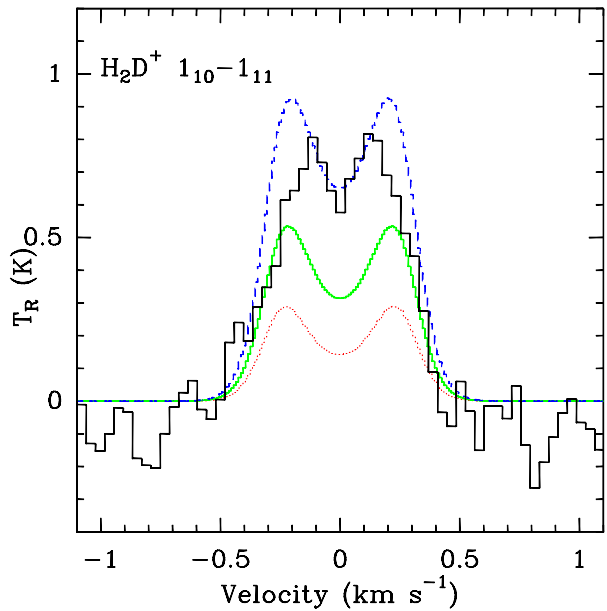}
\caption{Results of models with the temperature density distributions
  from Young et al, a `jump' in $b_D$ at $R$=10$^3$~AU, and with the
  \hhdp\ collisional rate coefficient at the standard value and
  increased by factors of 3 and 10 (curves from bottom to top),
  superposed on the observations (histogram).}
\label{f:bdj}
\end{figure}

Figure~\ref{f:bdj} shows the results of models with the CRC increased
by factors of 3 and 10 above the value from \citet{black90}. The line
strength is seen to scale with the CRC; this trend continues for the
models with the CRC decreased below `standard' which are not shown. 
The shape of the \hhdp\ profile is seen not to depend on the CRC: both
the emission from the core and the absorption in the outer layers
scale with the CRC.
For the adopted model parameters, the best fit to the observed \hhdp\
profile would be obtained by increasing the CRC by a factor of $\approx$5.
However, this number depends critically on several parameters,
particularly $n$(\hh), \tkin\ and the \hhdp\ abundance.
This is why we refrain from recommending a value for the CRC at this point.
Even though $n$(\hh) and \tkin\ can be obtained from other observations 
(as we have done), the \hhdp\ abundance cannot.
Therefore, for the time being, \hhdp\ abundances can only be
determined to factors of $\sim$10 accuracy. We recommend theoretical
calculations of the CRC of \hhdp\ to improve this situation.

The fact that the simplified kinematics give a better fit to the observed
\hhdp\ profile than the models in Section~\ref{s:alto} implies that the infall
motions toward \obj\ are very small, in good agreement with the conclusion
from \citet{williams:l1544} that the infall speed is much smaller than the
thermal, rotational and gravitational speeds in this source.
The observed intensity at the 20$''$ offset position of \tmb=0.4~K is well
reproduced by the same model that fits the central spectrum. The model predicts
a $\approx$20\% absorption at the line center, which the present data do not
rule out. However, better spectra at offset positions are needed to confirm this
prediction. 

\section{Conclusions}
\label{s:concl}

Radiative transfer modeling of line profiles of \hhdp, \hcop\ and
\nnhp\ in the pre-stellar core LDN~1544 shows that previous
descriptions of the temperature, density and velocity structure, that
reproduce \hcop\ and \nnhp\ data, do not fit \hhdp\ data, which probe
smaller radii.
At least two modifications are required at these radii, which affect
\hhdp, but not \hcop\ or \nnhp.

First, to make the excitation temperature of \hhdp\ increase inward,
the density cannot flatten off as proposed by \citet{ward-t99}, but
must continue to increase as in the models by \citet{evans01}. In
addition, only very small inward decreases of the gas temperature are
allowed. 
Observations of \ammo\ toward a sample of pre-stellar cores also
suggest a constant kinetic temperature with radius (Tafalla et al.\
\citeyear{tafalla02}, \citeyear{tafalla04}).
The same result is obtained from observations of CO isotopes toward
the pre-stellar core B~68 \citep{bergin05}, suggesting that current
models of pre-stellar cores either underestimate the dust temperature
because of changes in the dust opacity, or overestimate the gas-grain
thermal coupling because of grain growth.
However, the \ammo\ and CO observations probe larger radii than \hhdp,
and interferometric observations of \ammo\ would be valuable.

Second, the infall velocity needs to increase with radius down to
small radii. If cloud collapse is quasi-static and mediated by
ambipolar diffusion, the central region supported by thermal pressure,
where the infall speed drops to zero, must be smaller than in the
models by \citet{ciolek00}.  Alternatively, the collapse may be
`mildly' dynamic, indicated by values of 1--3 for the parameter $m_0$
\citep{hunter77,foster93}. Highly supersonic infall velocities, as
well as large-scale rotation, are ruled out.

Alternatively, the central dip on the \hhdp\ profile is due to absorption
  in the outer parts of the core, as seen for CS, HCO$^+$ and \nnhp. However,
  such absorption is not seen for DCO$^+$ 2--1, nor in \hhdp\ at offset
  positions. Also, to produce an absorption as narrow as observed, the outer
  layers of \obj\ must have essentially zero infall and turbulent motions,
  consistent with evidence from CS, HCO$^+$ and \nnhp\ \citep{williams:l1544}.
  To test this `absorption' hypothesis, sensitive mapping of the \hhdp\ line and
  of \dcop\ 1--0 are needed.

In the future, making full use of \hhdp\ as kinematic probe requires
maps at higher sensitivity and spatial and spectral resolution than
the CSO can offer.
Other single-dish telescopes (JCMT, APEX) can fulfill some of these
requirements, but not all.
Therefore, for a real breakthrough, interferometric observations will be
crucial.  The SMA will enable case studies, but only ALMA will have the
sensitivity to probe many pre-stellar cores within reasonable times.
The interpretation of the data will require good observational
constraints of the \textit{gas} temperature and density.
In addition, theoretical calculations of the
  collisional rate coefficients of \hhdp\ are urgently needed.

If the current observations cannot be matched with any of these
modifications, then we speculate that the doubled-peaked \hhdp\ line
profile in LDN~1544 is caused by the presence of two protostellar
condensations, orbiting each other, in the core center. We may be
witnessing the birth of a \textit{binary} protostellar system. Once
again, interferometer observations are required to test this
hypothesis.

\begin{acknowledgement}
  The authors thank Malcolm Walmsley, Arnaud Belloche, Neal Evans
  and the anonymous referee for useful comments on the
  manuscript, Glenn Ciolek and Kaisa Young for sending their model
  results in electronic form, and Aurore Bacmann and Antonio Crapsi
  for help with the observations. P.C.\ acknowledges support from the MIUR
  grant `Dust and molecules in astrophysical environments'.
\end{acknowledgement}

\bibliographystyle{aa}
\bibliography{h2d+}

\begin{thebibliography}{45}
\expandafter\ifx\csname natexlab\endcsname\relax\def\natexlab#1{#1}\fi

\bibitem[{{Aikawa} {et~al.}(2005){Aikawa}, {Herbst}, {Roberts}, \&
  {Caselli}}]{aikawa05}
{Aikawa}, Y., {Herbst}, E., {Roberts}, H., \& {Caselli}, P. 2005, \apj, 620,
  330

\bibitem[{{Andr{\'e}} {et~al.}(2000){Andr{\'e}}, {Ward-Thompson}, \&
  {Barsony}}]{andre:ppiv}
{Andr{\'e}}, P., {Ward-Thompson}, D., \& {Barsony}, M. 2000, Protostars and
  Planets IV, 59

\bibitem[{{Bacmann} {et~al.}(2003){Bacmann}, {Lefloch}, {Ceccarelli},
  {Steinacker}, {Castets}, \& {Loinard}}]{bacmann03}
{Bacmann}, A., {Lefloch}, B., {Ceccarelli}, C., {et~al.} 2003, \apjl, 585, L55

\bibitem[{{Belloche} \& {Andr{\' e}}(2004)}]{belloche04}
{Belloche}, A. \& {Andr{\' e}}, P. 2004, \aap, 419, L35

\bibitem[{{Belloche} {et~al.}(2002){Belloche}, {Andr{\' e}}, {Despois}, \&
  {Blinder}}]{arnaud02}
{Belloche}, A., {Andr{\' e}}, P., {Despois}, D., \& {Blinder}, S. 2002, \aap,
  393, 927

\bibitem[{{Bergin} {et~al.}(2002){Bergin}, {Alves}, {Huard}, \&
  {Lada}}]{bergin02}
{Bergin}, E.~A., {Alves}, J., {Huard}, T., \& {Lada}, C.~J. 2002, \apjl, 570,
  L101

\bibitem[{{Bergin} \& {Langer}(1997)}]{bergin97}
{Bergin}, E.~A. \& {Langer}, W.~D. 1997, \apj, 486, 316

\bibitem[{{Bergin} {et~al.}(2005){Bergin}, {Maret}, \& {van der
  Tak}}]{bergin05}
{Bergin}, E.~A., {Maret}, S., \& {van der Tak}, F.~F.~S. 2005, in The Dusty and
  Molecular Universe, ed. A. Wilson (ESA-SP 577), 185--190

\bibitem[{{Black} {et~al.}(1990){Black}, {van Dishoeck}, {Willner}, \&
  {Woods}}]{black90}
{Black}, J.~H., {van Dishoeck}, E.~F., {Willner}, S.~P., \& {Woods}, R.~C.
  1990, \apj, 358, 459

\bibitem[{{Caselli}(2003)}]{paola03}
{Caselli}, P. 2003, \apss, 285, 619

\bibitem[{{Caselli} {et~al.}(2003){Caselli}, {van der Tak}, {Ceccarelli}, \&
  {Bacmann}}]{h2d+03}
{Caselli}, P., {van der Tak}, F.~F.~S., {Ceccarelli}, C., \& {Bacmann}, A.
  2003, \aap, 403, L37

\bibitem[{{Caselli} {et~al.}(2002{\natexlab{a}}){Caselli}, {Walmsley},
  {Zucconi}, {Tafalla}, {Dore}, \& {Myers}}]{paola02a}
{Caselli}, P., {Walmsley}, C.~M., {Zucconi}, A., {et~al.} 2002{\natexlab{a}},
  \apj, 565, 331

\bibitem[{{Caselli} {et~al.}(2002{\natexlab{b}}){Caselli}, {Walmsley},
  {Zucconi}, {Tafalla}, {Dore}, \& {Myers}}]{paola02b}
---. 2002{\natexlab{b}}, \apj, 565, 344

\bibitem[{{Ceccarelli} \& {Dominik}(2005)}]{ceccarelli05}
{Ceccarelli}, C. \& {Dominik}, C. 2005, \aap, submitted

\bibitem[{{Ceccarelli} {et~al.}(2004){Ceccarelli}, {Dominik}, {Lefloch},
  {Caselli}, \& {Caux}}]{ceccarelli04}
{Ceccarelli}, C., {Dominik}, C., {Lefloch}, B., {Caselli}, P., \& {Caux}, E.
  2004, \apjl, 607, L51

\bibitem[{{Ciolek} \& {Basu}(2000)}]{ciolek00}
{Ciolek}, G.~E. \& {Basu}, S. 2000, \apj, 529, 925

\bibitem[{{Crapsi} {et~al.}(2004){Crapsi}, {Caselli}, {Walmsley}, {Tafalla},
  {Lee}, {Bourke}, \& {Myers}}]{crapsi04}
{Crapsi}, A., {Caselli}, P., {Walmsley}, C.~M., {et~al.} 2004, \aap, 420, 957

\bibitem[{{Evans} {et~al.}(2001){Evans}, {Rawlings}, {Shirley}, \&
  {Mundy}}]{evans01}
{Evans}, N.~J., {Rawlings}, J.~M.~C., {Shirley}, Y.~L., \& {Mundy}, L.~G. 2001,
  \apj, 557, 193

\bibitem[{{Foster} \& {Chevalier}(1993)}]{foster93}
{Foster}, P.~N. \& {Chevalier}, R.~A. 1993, \apj, 416, 303

\bibitem[{{Frerking} {et~al.}(1982){Frerking}, {Langer}, \&
  {Wilson}}]{frerking82}
{Frerking}, M.~A., {Langer}, W.~D., \& {Wilson}, R.~W. 1982, \apj, 262, 590

\bibitem[{{Galli} {et~al.}(2002){Galli}, {Walmsley}, \& {Gon{\c
  c}alves}}]{galli02}
{Galli}, D., {Walmsley}, M., \& {Gon{\c c}alves}, J. 2002, \aap, 394, 275

\bibitem[{{Hartmann} {et~al.}(2001){Hartmann}, {Ballesteros-Paredes}, \&
  {Bergin}}]{hartmann01}
{Hartmann}, L., {Ballesteros-Paredes}, J., \& {Bergin}, E.~A. 2001, \apj, 562,
  852

\bibitem[{{Hogerheijde}(2001)}]{michiel01}
{Hogerheijde}, M.~R. 2001, \apj, 553, 618

\bibitem[{{Hogerheijde} \& {van der Tak}(2000)}]{hst}
{Hogerheijde}, M.~R. \& {van der Tak}, F.~F.~S. 2000, \aap, 362, 697

\bibitem[{{Hunter}(1977)}]{hunter77}
{Hunter}, C. 1977, \apj, 218, 834

\bibitem[{{Mundy} {et~al.}(2000){Mundy}, {Looney}, \& {Welch}}]{mundy00}
{Mundy}, L.~G., {Looney}, L.~W., \& {Welch}, W.~J. 2000, Protostars and Planets
  IV, 355

\bibitem[{{Myers}(2005)}]{myers05}
{Myers}, P.~C. 2005, \apj, in press; astro-ph/0501127

\bibitem[{{Pagani} {et~al.}(1992){Pagani}, {Salez}, \& {Wannier}}]{pagani:h2d+}
{Pagani}, L., {Salez}, M., \& {Wannier}, P.~G. 1992, \aap, 258, 479

\bibitem[{{Roberts} {et~al.}(2003){Roberts}, {Herbst}, \& {Millar}}]{roberts03}
{Roberts}, H., {Herbst}, E., \& {Millar}, T.~J. 2003, \apjl, 591, L41

\bibitem[{{Rodgers} \& {Charnley}(2003)}]{rodgers:cocoon}
{Rodgers}, S.~D. \& {Charnley}, S.~B. 2003, \apj, 585, 355

\bibitem[{{Schmid-Burgk} {et~al.}(2004){Schmid-Burgk}, {Muders}, {M{\" u}ller},
  \& {Brupbacher-Gatehouse}}]{jsb04}
{Schmid-Burgk}, J., {Muders}, D., {M{\" u}ller}, H.~S.~P., \&
  {Brupbacher-Gatehouse}, B. 2004, \aap, 419, 949

\bibitem[{{Sch\"{o}ier} {et~al.}(2005){Sch\"{o}ier}, {van der Tak}, {van
  Dishoeck}, \& Black}]{moldata}
{Sch\"{o}ier}, F.~L., {van der Tak}, F.~F.~S., {van Dishoeck}, E.~F., \& Black,
  J.~H. 2005, \aap, 432, 369

\bibitem[{{Shen} {et~al.}(2004){Shen}, {Greenberg}, {Schutte}, \& {van
  Dishoeck}}]{shen04}
{Shen}, C.~J., {Greenberg}, J.~M., {Schutte}, W.~A., \& {van Dishoeck}, E.~F.
  2004, \aap, 415, 203

\bibitem[{{Shu} {et~al.}(2000){Shu}, {Najita}, {Shang}, \& {Li}}]{shu:ppiv}
{Shu}, F.~H., {Najita}, J.~R., {Shang}, H., \& {Li}, Z.-Y. 2000, Protostars and
  Planets IV, 789

\bibitem[{{Tafalla} {et~al.}(1998){Tafalla}, {Mardones}, {Myers}, {Caselli},
  {Bachiller}, \& {Benson}}]{tafalla98}
{Tafalla}, M., {Mardones}, D., {Myers}, P.~C., {et~al.} 1998, \apj, 504, 900

\bibitem[{{Tafalla} {et~al.}(2004){Tafalla}, {Myers}, {Caselli}, \&
  {Walmsley}}]{tafalla04}
{Tafalla}, M., {Myers}, P.~C., {Caselli}, P., \& {Walmsley}, C.~M. 2004, \aap,
  416, 191

\bibitem[{{Tafalla} {et~al.}(2002){Tafalla}, {Myers}, {Caselli}, {Walmsley}, \&
  {Comito}}]{tafalla02}
{Tafalla}, M., {Myers}, P.~C., {Caselli}, P., {Walmsley}, C.~M., \& {Comito},
  C. 2002, \apj, 569, 815

\bibitem[{{Van der Tak} {et~al.}(2004){Van der Tak}, {Caselli}, {Walmsley},
  {Bacmann}, \& {Crapsi}}]{zermatt}
{Van der Tak}, F.~F.~S., {Caselli}, P., {Walmsley}, C.~M.~{Ceccarelli}, C.,
  {Bacmann}, A., \& {Crapsi}, A. 2004, in The Dense Interstellar Medium in
  Galaxies (Springer-Verlag), 549--552

\bibitem[{{Van der Tak} \& {van Dishoeck}(2000)}]{zeta}
{Van der Tak}, F.~F.~S. \& {van Dishoeck}, E.~F. 2000, \aap, 358, L79

\bibitem[{{Vastel} {et~al.}(2004){Vastel}, {Phillips}, \& {Yoshida}}]{vastel04}
{Vastel}, C., {Phillips}, T.~G., \& {Yoshida}, H. 2004, \apjl, 606, L127

\bibitem[{{Walmsley} {et~al.}(2004){Walmsley}, {Flower}, \& {Pineau des For{\^
  e}ts}}]{walmsley04}
{Walmsley}, C.~M., {Flower}, D.~R., \& {Pineau des For{\^ e}ts}, G. 2004, \aap,
  418, 1035

\bibitem[{{Ward-Thompson} {et~al.}(1999){Ward-Thompson}, {Motte}, \&
  {Andr{\'e}}}]{ward-t99}
{Ward-Thompson}, D., {Motte}, F., \& {Andr{\'e}}, P. 1999, \mnras, 305, 143

\bibitem[{{Williams} {et~al.}(1999){Williams}, {Myers}, {Wilner}, \& {di
  Francesco}}]{williams:l1544}
{Williams}, J.~P., {Myers}, P.~C., {Wilner}, D.~J., \& {di Francesco}, J. 1999,
  \apjl, 513, L61

\bibitem[{{Young} {et~al.}(2004){Young}, {Lee}, {Evans}, {Goldsmith}, \&
  {Doty}}]{young:h2co}
{Young}, K.~E., {Lee}, J., {Evans}, N.~J., {Goldsmith}, P.~F., \& {Doty}, S.~D.
  2004, \apj, 614, 252

\bibitem[{{Zhou}(1992)}]{zhou92}
{Zhou}, S. 1992, \apj, 394, 204

\end{thebibliography}

\end{document}